\documentclass[twocolumn,pra,aps]{revtex4}
\usepackage{graphicx}

\def\duzomniejsze{<\kern-.7mm<}
\def\duzowieksze{>\kern-.7mm>}

\def\textbf#1{{\bf #1}}
\def\beq{\begin{equation}}
\def\eeq{\end{equation}}
\def\be{\begin{equation}}
\def\ee{\end{equation}}
\def\ben{\begin{eqnarray}}
\def\een{\end{eqnarray}}
\def\beqa{\begin{eqnarray}}
\def\eeqa{\end{eqnarray}}
\def\eea{\end{array}}
\def\bea{\begin{array}}
\newcommand{\bei}{\begin{itemize}}
\newcommand{\eei}{\end{itemize}}
\newcommand{\bee}{\begin{enumerate}}
\newcommand{\eee}{\end{enumerate}}

\def\>{\rangle}
\def\<{\langle}

\begin{document}

\title{Diffusion and entanglement of a kicked particle in an infinite square well under frequent measurements}

\begin{abstract}
We investigate the dynamics of a kicked particle in an infinite
square well undergoing frequent measurements of energy. For a
large class of periodic kicking forces, constant diffusion is
found in such a non-KAM system. The influence of phase shift of
the kicking potential on the short-time dynamical behavior is
discussed. The general asymptotical measurement-assisted diffusion
rate is obtained. The entanglement between the particle and the
measuring apparatus is investigated. There exist two distinct
dynamical behaviors of entanglement. The bipartite entanglement
between the system of interest and the whole spins of the
measuring apparatus grows with the kicking steps and it gains
larger value for the more chaotic system. However, the partial
entanglement between the system of interest and the partial spins
of the measuring apparatus decreases with the kicking steps. The
relation between the entanglement and quantum diffusion is also
analyzed.
\\

\end{abstract}
\author{Shang-Bin Li}\email{stephenli74@yahoo.com.cn}, \author{Jing-Bo Xu}

\affiliation{Chinese Center of Advanced Science and Technology
(World Laboratory), P.O.Box 8730, Beijing, People's Republic of
China;} \affiliation{Zhejiang Institute of Modern Physics and
Department of Physics, Zhejiang University, Hangzhou 310027,
People's Republic of China}

\maketitle

\section * {I. INTRODUCTION}

One of the significant consequences of quantum mechanics is that
the measurement unavoidably disturbs the measured system. This is
particularly revealed by the so-called Zeno and anti-Zeno effects
\cite{Misra1977,Kofman1996,Kofman2000}. The Zeno (anti-Zeno)
effect refers to the inhibition (acceleration) of the evolution
when one attempts to observe it, which can be regarded as two
particular consequences of the disturbance of the observed system
caused by quantum measurement. The first experiment on the
quantum-Zeno effect was proposed by Cook \cite{Cook1988}, and
realized by Itano et al. \cite{Itano1990} using coherent Rabi
oscillations in a three-level atom. An opposite phenomenon, the
anti-Zeno effect (decay acceleration by frequent measurements) was
recently discovered by Kofman and Kurizki
\cite{Kofman1996,Kofman2000}. In the theoretical and numerical
investigations of quantum chaotic systems, similar predictions have been presented \cite{Kaulakys1997,Antoniou2001,Facchi1999,Facchi2001,Fischer2001}. \\

Dynamical localization, i.e. the quantum mechanical suppression of
classical chaotic diffusion, was firstly discovered by Casati et
al. \cite{Casati1979} in their investigation of the kicked rotator
(standard map), which can be understood as a dynamical version of
Anderson localization \cite{Fishman1982}. Quantum localization
emerges due to quantum interference, which can be destroyed by
noise and interactions with the environment, i.e. the decoherence.
Many theoretical and experimental studies have shown that even a
small amount of noise demolishes localization
\cite{Arnmann1998,Oberthaler1999,Shepelyansky1983,Ott1984}.
Quantum measurements can be regarded as another type of coupling
to the "environment", i.e. the measurement apparatus
\cite{Zurek2003}. Kaulakys and Gontis have shown that, in the case
of the kicked rotator, a diffusive behavior can emerge even in the
quantum case, if a projective measurement is performed after every
kick \cite{Kaulakys1997}. Facchi et al. have found that projective
measurements provoke diffusion even when the corresponding
classical dynamics is regular \cite{Facchi1999}. Dittrich and
Graham have studied the kicked rotor coupled to macroscopic
systems acting as continuous measuring devices with limited time
resolution and found the diffusive energy growth
\cite{Dittrich1990}. Most previous works concerning the anti-Zeno
effect in quantum chaos have been concentrated on quantum systems
whose classical counterparts obey the Kolmogorov-Arnold-Moser
(KAM) theorem. So, it is interesting to investigate the influence
of frequent measurements on the dynamical behavior of the non-KAM
system, such as a kicked particle in an infinite square well
\cite{Hu1999}. In this paper, we investigate the dynamics of a
kicked particle in an infinite square well undergoing frequent
measurements of energy. It is found that, for large class of
periodic kicking forces, the dynamical behaviors exhibit diffusion
with constant rate in such a non-KAM system. Then, we investigate
how the ratio of the well width and the kicking field wavelength
affect the diffusion of energy in the present situations. It is
shown that not only increasing the kicking field strength, but
also increasing the ratio of the well width and the kicking field
wavelength can enhance the diffusion of energy in such a non-KAM
system undergoing repeated measurements.

Recently, much attention has also been paid to the entanglement in
the quantum chaotic systems. Several authors have studied the
entanglement in coupled kicked tops or Dicke model
\cite{Furuya1998,Miller1999,Lakshminarayan2001,Bandyopadhyay2002,Tanaka2002,Emary2003,Fujisaki2003,Bandyopadhyay2004,Hou2004,Wang2004,Rossini2004,Demkowicz-Dobrzanski2004,Ghose2004},
and found it has a manifestation of chaotic behavior. It has been
demonstrated that classical chaos can lead to substantially
enhanced entanglement and it has been shown that entanglement
provides a useful characterization of quantum states in
higher-dimensional chaotic systems. For the system of coupled
kicked tops, it has been clarified that two initially separable
subsystems can get entangled in a nearly linear rate depending on
the intrinsic chaotic properties, and their entanglement
eventually reach a saturation
\cite{Miller1999,Lakshminarayan2001,Bandyopadhyay2002,Tanaka2002,Fujisaki2003,Bandyopadhyay2004}.
It has also been elucidated that the increment of the nonlinear
parameter of coupled kicked tops does not accelerate the
entanglement production in the strongly chaotic region
\cite{Fujisaki2003}. For a single kicked top composed of
collective spin-$\frac{1}{2}$, it has been suggested that
entanglement can be regarded as a signature of quantum chaos. Both
bipartite entanglement and pairwise entanglement in the spins have
been considered and it has been revealed that bipartite
entanglement is enhanced in the chaotic region, nevertheless
pairwise entanglement is suppressed \cite{Wang2004}. Most of the
previous works studied how the intrinsic dynamical properties of
the quantum chaotic systems affect the entanglement between the
subsystems. It may be more natural to explore how the entanglement
between the system of interest and its surrounding environment or
measuring apparatus affects the chaotic behavior such as the
diffusion behavior. Here, we show that the diffusion of the kicked
particle in an infinite square well undergoing frequent
measurements of energy is closely related with the entanglement of
the particle and the measuring apparatus. It is found that there
exist two distinct dynamical behaviors of entanglement. The
bipartite entanglement (defined as entanglement between the
particle and the whole degree of freedom of the measuring
apparatus in this paper) grows with the kicking steps and it gains
larger value for the more chaotic system. However, the partial
entanglement between pairs (specifically defined as entanglement
between the particle and the partial degree of freedom of the
measuring apparatus in this paper) decreases with the kicking
steps. It is very desirable to investigate the asymptotical
behavior of the bipartite entanglement or the partial entanglement
between pairs. Can the bipartite entanglement reach a saturation?
In this paper, these are still the open questions.

This paper is organized as follows: In Sec.II, we investigate the
diffusion in the non-KAM systems. The dynamics of a kicked
particle in an infinite square well undergoing frequent
measurements of energy is studied in details. It is found that,
for large class of periodic kicking forces, the dynamical
behaviors exhibit diffusion with constant rate in such a non-KAM
system. Then, we investigate how the ratio of the well width and
the kicking field wavelength affect the diffusion of energy in the
present situations. It is shown that not only increasing the
kicking field strength, but also increasing the ratio of the well
width and the kicking field wavelength can enhance the diffusion
of energy in such a non-KAM system undergoing repeated
measurements. In Sec.III, we focus our attention on the
entanglement between this quantum chaotic system and the measuring
apparatus by exploring the relative entropy of entanglement. We
investigate how the inherent quantum chaos affects the
entanglement between the particle and measuring apparatus. Two
distinct dynamical behaviors of entanglement are revealed. The
bipartite entanglement grows with the kicking steps. It increases
with a higher rate for the more chaotic system in the short time.
However, the partial entanglement between the particle and the
partial spins of the measuring apparatus decreases with the
kicking steps. Some conclusive remarks and brief discussion about
the experimental verification of measurement-induced diffusion in
the quantum dot are given in Sec.IV.

\section * {II. DIFFUSION OF THE KICKED PARTICLE IN AN INFINITE SQUARE WELL UNDERGOING FREQUENT MEASUREMENTS OF ENERGY}

In this section, we investigate the diffusion of the kicked
particle in an infinite square well undergoing frequent
measurements of energy. The Hamiltonian investigated here is
described by \be
H=\frac{p^2}{2}+V_0(x)+V(x)\sum^{\infty}_{l=-\infty}\delta(t-lT),
\ee where the potential $V_0(x)$ is the confining infinite square
well potential of width $\pi$, centered at the position
$\frac{\pi}{2}$. $V(x)$ is the external potential. Since the two
hard walls destroy the analyticity of the potential, the KAM
scenario break down in the system (1). In Ref.\cite{Hu1999}, the
authors have studied quantum chaos of the system (1) with
$V(x)=k\cos(x+\alpha)$, where $\alpha$ is a phase shift. It was
shown that, for a small perturbation $K$($=kT$), the classical
phase space displays a stochastic web structure, and the diffusion
rate scales as $D\propto{K}^{2.5}$. However, in the large $K$
regime, $D\propto{K}^{2}$. Quantum mechanically, they observed
that the quasi-eigenstates are power-law localized for small $K$
and extended for large $K$. In what follows, we investigate the
evolution of system (1) interrupted by quantum measurements, in
the following sense: the system evolves under the action of the
free Hamiltonian $\frac{P^2}{2}+V_0(x)$ for
$(N-1)T+t^{\prime}<t<NT$ ($0<t^{\prime}<T$), undergoes a kick at
$t=NT$, evolves again freely. Then a measurement of the energy $E$
is acted on the system at $t=NT+t^{\prime}$. The evolution of the
density matrix can be written as follows \beqa
&&\rho_{NT+t^{\prime}}={\mathcal{L}}^N\rho_{t^{\prime}},\nonumber\\
&&{\mathcal{L}}\rho=\sum^{\infty}_{n=1}U_{free}(t^{\prime})U_{kick}U_{free}(T-t^{\prime})
|n\rangle\langle{n}|\rho|n\rangle\langle{n}|\nonumber\\
&&~~~~~~~~~~~~\cdot{U}^{\dagger}_{free}(T-t^{\prime})U^{\dagger}_{kick}U^{\dagger}_{free}(t^{\prime}),
\eeqa where $\rho_{NT+t^{\prime}}$ represents the density matrix
of the particle at the time $NT+t^{\prime}$, $|n\rangle$ is the
$n$th eigenstate of the nonperturbed Hamiltonian, and \beqa
&&U_{kick}=\exp[-iV(x)/\hbar],~~~U_{free}(t)=\exp(-i\frac{p^2t}{2\hbar}),\nonumber\\
&&\langle{x}|n\rangle=\sqrt{\frac{2}{\pi}}\sin(nx),~x\in[0,\pi],~n=1,2...
\eeqa From Eq.(2), we can derive the occupation probabilities
$P_n(N)\equiv\langle{n}|\rho_{NT+t^{\prime}}|n\rangle$ which are
governed by \be P_{n}(N)=\sum_{m}Z_{nm}P_m(N-1),\ee where \be
Z_{nm}=|\langle{n}|U_{kick}|m\rangle|^2 \ee are the transition
probabilities. In Fig.1, we display the evolution of occupation
probabilities governed by Eqs.(4-5) with the kicking potential
$V(x)=k\cos(x+1)$ for different values of $k$. It is clearly shown
that the particle can be driven by the external periodic field
from the ground level to the higher excited levels.
\begin{figure}
\includegraphics{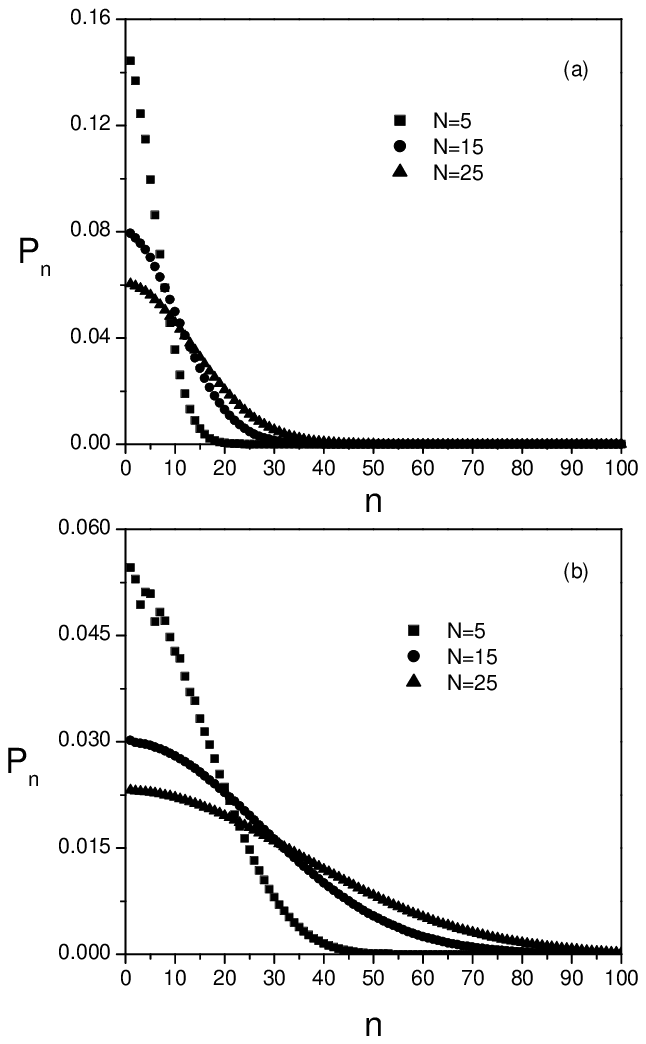}
\caption{ The occupation probabilities $P_n$ of the particle
undergoing the $N$th kick and the $N$th projective measurement of
the energy are depicted for different values of $k$. The particle
is initially in the ground state and the kicking potential
$V(x)=k\cos(x+1)$ is chosen as an illustration. (a) $k=4\hbar$,
(b) $k=10\hbar$.}
\end{figure}
\begin{figure}
\includegraphics{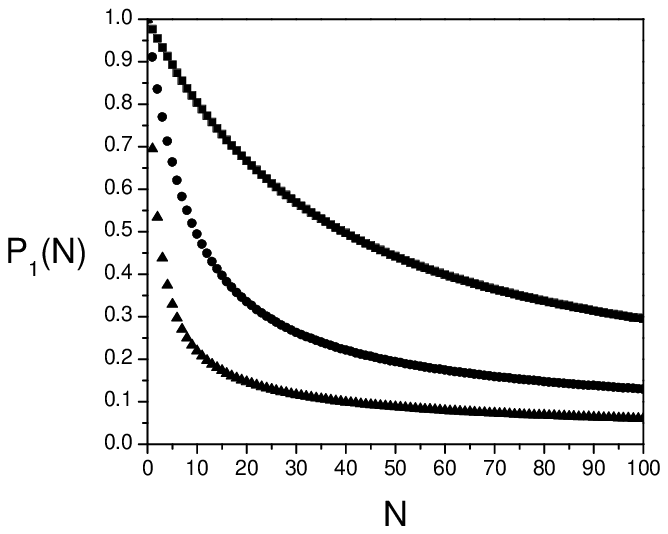}
\caption{The ground occupation probabilities $P_1$ of the particle
at the time $NT+t^{\prime}$ are plotted as a function of $N$ for
different values of $k$. The particle is initially in the ground
state and the kicking potential $V(x)=k\cos(x+1)$ is chosen as an
illustration. (Solid Square) $k=0.5\hbar$, (Solid Circle)
$k=\hbar$, (Solid Triangle) $k=2\hbar$.}
\end{figure}
In what follows, we will derive an analytical expression for the
diffusion rate defined by $D\equiv\frac{E_N-E_0}{N}$, where $E_N$
is the expected value of energy of the particle at the time
$NT+t^{\prime}$. By making use of the Eqs.(4-5), we can obtain
\beqa
E_N&=&\sum^{\infty}_{n=1}\frac{\hbar^2n^2}{2}P_n(N)\nonumber\\
&=&E_{N-1}+\frac{1}{2\pi}\int^{\pi}_{0}(V^{\prime})^2dx\nonumber\\
&&-\frac{1}{2\pi}\sum^{\infty}_{m=1}\int^{\pi}_{0}(V^{\prime})^2\cos(2mx)dxP_{m}(N-1),
\eeqa where $V^{\prime}$ denotes the first-order derivative of
$V(x)$ upon $x$. From Eq.(6), we can derive a sufficient condition
related to the diffusion with constant rate. If $V(x)$ satisfies
the following expression: \be
(V^{\prime})^2=a_0+\sum^{\infty}_{m=1}a_{m}\sin(2mx), \ee where
$a_{m}$ ($m=0,1,2,...$) are the expansion coefficients which
should ensure the positive definite of the function
$a_0+\sum^{\infty}_{m=1}a_{m}\sin(2mx)$, the diffusion rate $D$ is
the constant $\frac{a_0}{2}$. As an illustration, we consider the
case with $V(x)=k\cos(x+\frac{\pi}{4})$, which obviously satisfies
the Eq.(7) and the corresponding expansion coefficients are
$a_0=a_1=\frac{k^2}{2}$ ($a_m=0$ for $m>1$). Then, in this case,
iterating the Eq.(6), we can obtain \be E_N=E_0+\frac{k^2N}{4},
\ee which leads to $D=\frac{k^2}{4}$. For $V(x)=k\cos(x+\alpha)$,
we have \be
E_N=E_{N-1}+\frac{k^2}{4}+\frac{k^2\cos(2\alpha)}{8}P_1(N-1). \ee
The above equation shows that the increase of energy after every
step is closely related to the ground state population of the
previous step. Eq.(9) also implies that the phase shift $\alpha$
plays a significant role in the short-time dynamical behavior of
the kicked particle in this situation. The diffusion will be
enhanced or suppressed by adjusting the sign of $\cos(2\alpha)$.
However, since $\lim_{N\rightarrow\infty}P_1(N)=0$, the energy
diffusion is asymptotically linear and independent of $\alpha$ for
any values of $k$. In Fig.(2), the values of $P_1$ at the time
$NT+t^{\prime}$ are plotted as the function of $N$ for different
values of $k$. It is shown that $P_1(N)$ decreases with $N$ and,
the larger the value of $k$, the more rapid the decay of the
particle from the ground state. Therefore, according to the
Eq.(9), we know that the corresponding diffusion rate of the
non-unitary evolution governed by Eq.(2) with $V(x)=k\cos(x+1)$
increases with $N$ and eventually converges to $\frac{k^2}{4}$. \\

In Ref.\cite{Sankaranarayanan2001}, the authors have investigated
the quantum chaotic dynamics of the system (1) with
$V(x)=\cos(2Rx)$, where $R$ is a ratio of two length scales,
namely, the well width and the kicking field wavelength. If $R$ is
a noninteger the dynamics is non-KAM. It was shown that time
evolving states exhibit considerable $R$ dependence, and tuning
$R$ to enhance classical diffusion can lead to significantly
larger quantum diffusion for the same field strengths. It is
interesting to investigate how the values of $R$ affect the
diffusion of energy in the present situations. Substituting
$V(x)=k\cos(2Rx)$ into Eq.(6), we can obtain \beqa
E_N&=&E_{N-1}+k^2R^2-\frac{k^2R\sin(4R\pi)}{4\pi}\nonumber\\
&&+\frac{k^2R^3\sin(4R\pi)}{\pi}\sum^{\infty}_{m=1}\frac{1}{4R^2-m^2}P_m(N-1).\nonumber\\
\eeqa The above equation implies that not only increasing the
kicking field strength $k$, but also increasing the ratio $R$ of
the well width and the kicking field wavelength can enhance the
diffusion of energy in such a non-KAM system undergoing repeated
measurements. For elucidating it, we merely need to consider three
different cases listed in the following: (1) When $R$ is a integer
or half integer (i.e. $\frac{1}{2}$, $\frac{3}{2}$,
$\frac{5}{2}$...) , the Eq.(10) can be replaced by \be
E_N=E_{N-1}+k^2R^2+\frac{k^2R^2}{2}P_{2R}(N-1), \ee which is very
similar with Eq.(9). (2) When $R=\frac{2j+1}{4}$ ($j=0,1,2...$),
the Eq.(10) can be rewritten as \be E_N=E_{N-1}+k^2R^2, \ee which
implies the diffusion rate is a constant $k^2R^2$. (3) For any
other values of $R$, the diffusion behavior becomes very complex.
In what follows, we analyze both the short-time dynamical behavior
and the asymptotical dynamical behavior. For $V(x)=k\cos(2Rx)$,
the transition matrix $\langle{n}|U_{kick}|m\rangle$ can be
expressed as \beqa
\langle{n}|U_{kick}|m\rangle&=&J_0(k/\hbar)(\delta_{n,m}-\delta_{n,-m})\nonumber\\
&&+\frac{1}{\pi}\sum^{\infty}_{j=1}(-i)^jJ_j(k/\hbar)\nonumber\\
&&\cdot[C_j(n-m)-C_j(n+m)], \eeqa where $J_n(k/\hbar)$ is the
Bessel function and \beqa C_j(l)=
 \left\{\begin{array}{c}
 \frac{4(-1)^ljR\sin(2jR\pi)}{4j^2R^2-l^2}~~~{\mathrm{for}}~2jR\neq|l| \\\\
    ~~~\pi ~~~~~~~~~~~~~~~~~~~~~{\mathrm{for}}~2jR=|l|.
  \end{array}\right\}
 \eeqa
In Fig.(3), the decay from the initial ground state of the
particle is depicted for different values of $R$. It is shown
that, the larger the value of $R$, the more rapid the decay of the
particle from the ground state.
\begin{figure}
\includegraphics{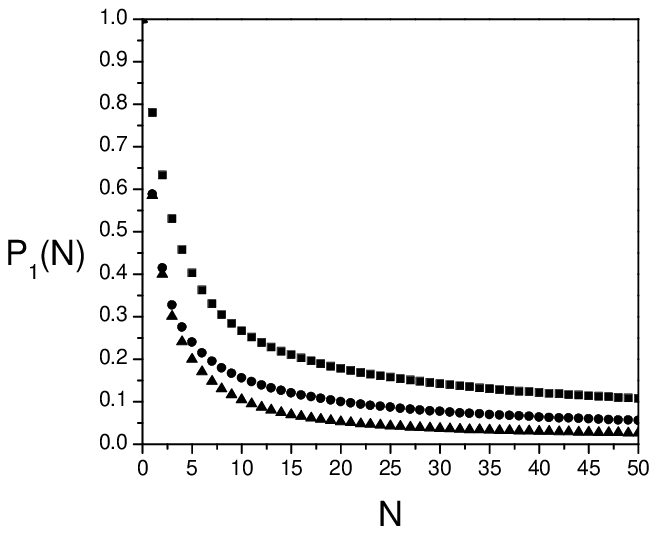}
\caption{ The ground occupation probabilities $P_1$ of the
particle at the time $NT+t^{\prime}$ are plotted as a function of
$N$ for different values of $R$. The particle is initially in the
ground state and the kicking potential $V(x)=k\cos(2Rx)$ is chosen
as an illustration, $k=\hbar$. (Solid Square) $R=\pi/4$, (Solid
Circle) $R=\pi/2$, (Solid Triangle) $R=\pi$.}
\end{figure}

If the particle is initially in the ground state, i.e.,
$P_n(0)=\delta_{n,1}$, we have $P_m(N-1)=(Z^{N-1})_{m,1}$, where
$Z\equiv{Z_{l_1,l_2}}$ is the transition probabilities matrix
defined by Eq.(5). Substituting it into the Eq.(10), and iterating
it, we can obtain \beqa
E_N&=&E_{0}+Nk^2R^2-\frac{Nk^2R\sin(4R\pi)}{4\pi}\nonumber\\
&&+\frac{k^2R^3\sin(4R\pi)}{\pi}\sum^{N-1}_{i=0}\sum^{\infty}_{m=1}\frac{1}{4R^2-m^2}(Z^i)_{m,1}\nonumber\\
&=&E_{0}+Nk^2R^2-\frac{Nk^2R\sin(4R\pi)}{4\pi}\nonumber\\
&&+\frac{k^2R^3\sin(4R\pi)}{\pi}\sum^{\infty}_{m=1}\frac{1}{4R^2-m^2}\left(\frac{1-Z^N}{1-Z}\right)_{m,1}.\nonumber\\
\eeqa In deriving the Eq.(15), we have used the formulation \be
\sum^{N-1}_{i=0}(Z^i)_{m,1}=\left(\frac{1-Z^N}{1-Z}\right)_{m,1}.
\ee It is not difficult to verify that \be
\lim_{N\rightarrow\infty}\left(\frac{1-Z^N}{1-Z}\right)_{m,1}=\left(\frac{1}{1-Z}\right)_{m,1},
\ee which implies that \be
\lim_{N\rightarrow\infty}\frac{E_N-E_0}{N}=k^2R^2-\frac{k^2R\sin(4R\pi)}{4\pi}.
\ee The right side of Eq.(18) is a monotonous increasing function
of $R$. So, one can always enhance the asymptotical diffusion rate
by increasing the ratio of the well width and the kicking field
wavelength in this case. In fact, the above deriving procedure can
also be applied to the Eq.(6), and the asymptotical diffusion rate
$\lim_{N\rightarrow\infty}\frac{E_N-E_0}{N}$ for any kicking
potential $V(x)$ is given by
$\frac{1}{2\pi}\int^{\pi}_{0}(V^{\prime})^2dx$ (Note that the
potential $V(x)$ is assumed to be differentiable). For clarifying
it , we substitute $P_m(N-1)=(Z^{N-1})_{m,1}$ into Eq.(6), where
the particle is assumed to be in the ground state. Similarly,
after iterating it, we have \beqa E_N
&=&E_{0}+\frac{N}{2\pi}\int^{\pi}_{0}(V^{\prime})^2dx\nonumber\\
&&-\frac{1}{2\pi}\sum^{\infty}_{m=1}\int^{\pi}_{0}(V^{\prime})^2\cos(2mx)dx\left(\frac{1-Z^N}{1-Z}\right)_{m,1}.
\eeqa Since the third term of the right side of Eq.(19) converges
to a finite value as $N\rightarrow\infty$ (Here, without loss of
generality, it is assumed that $\frac{1}{1-Z}$ is well-defined),
the asymptotical diffusion rate is given by
$\frac{1}{2\pi}\int^{\pi}_{0}(V^{\prime})^2dx$, which is very
similar with the result obtained in Ref.\cite{Facchi1999} for the
measurement-induced diffusion of the kicked rotor. This similarity
reveals the asymptotical equivalence caused by frequent
measurements of the kicked rotor and the kicked particle confined
in a infinite well, although the former is a KAM system and the
latter is a non-KAM system.

\section * {III. ENTANGLEMENT BETWEEN THE PARTICLE AND THE MEASURING APPARATUS}

As mentioned in Ref.\cite{Hu1999}, the experimental realization of
system (1) can be achieved by putting cold atoms in a quasi-1D
quantum dot. The atoms are then driven by a periodically pulsed
standing wave of light. Then, similar to the procedure discussed
in Ref.\cite{Facchi1999}, the projective measurement of the energy
at the time $t^{\prime}$ after every kick can be schematized by
associating an additional degree of freedom, such as a spin, with
every energy eigenstate. This is easily achieved by adding the
following Hamiltonian \cite{Pascazio1994} to system (1), \be
H_{mea}=\frac{\pi\hbar}{2}\sum_{n,N}|n\rangle\langle{n}|\otimes\sigma^{(n,N)}_1\delta(t-NT-t^{\prime}),
\ee where
$\sigma^{(n,N)}_1=|+\rangle_{(n,N)}\langle-|+|-\rangle_{(n,N)}\langle+|$
is the first Pauli matrix of the spin recording the energy
information in channel $(n,N)$, where $|+\rangle_{(n,N)}$,
$|-\rangle_{(n,N)}$ denote spin up state and spin down state of
the spin $\sigma^{(n,N)}$, respectively. In fact, the evolution
operator $U(NT+t^{\prime+},NT+t^{\prime-})$ caused by $H_{mea}$
can be simply expressed as \beqa
U(NT+t^{\prime+},NT+t^{\prime-})&=&\exp[-i\frac{\pi}{2}\sum_{n}|n\rangle\langle{n}|\otimes\sigma^{(n,N)}_1]\nonumber\\
&=&-i\sum_{n}|n\rangle\langle{n}|\otimes\sigma^{(n,N)}_1, \eeqa
which is actually a generalized d-dimensional control-NOT gate
operation. Control-NOT gate operation can be regarded as an
entangler. We assume that all of the spins are initially in the
spin down states. It is not difficult to demonstrate that the
total Hamiltonian of (1) plus $H_{mea}$ can result in the
evolution governed by Eq.(2) of the reduced density matrix of the
measured system. In order to verify it, we need only insert
$U[(N-1)T+t^{\prime+},(N-1)T+t^{\prime-}]$ into the time evolution
operator generated by the Hamiltonian (1) \beqa
\rho_{NT+t^{\prime-}}&=&{\mathrm{Tr}}_{\{\sigma^{(N-1)}\}}[
\sum_{n,m}
U_1|n\rangle\langle{n}|\rho_{(N-1)T+t^{\prime-}}|m\rangle\langle{m}|U^{\dagger}_1\nonumber\\
&&\otimes\sigma^{(n,N-1)}_1\bigotimes_{j}|-\rangle_{(j,N-1)}\langle-|\sigma^{(m,N-1)}_1],\nonumber\\
&=&\sum_{n}U_1|n\rangle\langle{n}|\rho_{(N-1)T+t^{\prime-}}|n\rangle\langle{n}|U^{\dagger}_1\nonumber\\
\eeqa where
$U_1=U_{free}(t^{\prime})U_{kick}U_{free}(T-t^{\prime})$, and
$\rho_{NT+t^{\prime-}}$ is the reduced density matrix of the
particle at the time $NT+t^{\prime-}$. In Eq.(22),
${\mathrm{Tr}}_{\{\sigma^{(N-1)}\}}$ denotes the tracing over all
of the spins labelled by $\sigma^{(i,N-1)}$ ($i=1,2,....$).
Eq.(22) implies that the coupling between the system of interest
and the apparatus causes that the total system becomes an
entangled state and the reduced density matrix of the particle is
completely projected on the energy eigenstates. Iterating the
Eq.(22), we obtain \beqa
\rho_{NT+t^{\prime-}}&=&{\tilde{\mathcal{L}}}^N\rho_{t^{\prime-}},\nonumber\\
{\tilde{\mathcal{L}}}\rho&=&\sum_{n}U_1|n\rangle\langle{n}|\rho|n\rangle\langle{n}|{U}^{\dagger}_1,
\eeqa which is exactly the expression of Eq.(2). As mentioned
above, the system of interest and the measuring apparatus become
an entangled state in this situation. It should be very
interesting to investigate how the inherent quantum chaos affects
the entanglement between the particle and measuring apparatus. In
the following, we confine our discussions in the bipartite
entanglement between the particle and all of the spins, and the
partial entanglement between the particle and the partial spins
$\sigma^{(N)}$. It is found that the bipartite entanglement and
the partial entanglement between pairs exhibit two distinct
dynamical behaviors. For characterizing the entanglement, we adopt
the relative entropy of entanglement defined by
$E_r(\rho)=\min_{\varrho\in\Omega}S(\rho\parallel\varrho)$
\cite{Vedral1998}, where $\Omega$ is the set of all disentangled
states, and
$S(\rho\parallel\varrho)\equiv{\mathrm{Tr}}[\rho(\log_2\rho-\log_2\varrho)]$
is the quantum relative entropy. The relative entropy of
entanglement is a good measure of quantum entanglement, and it
reduces to the von-Neumann entropy of the reduced density matrix
of either subsystems for pure states. So the bipartite
entanglement between the particle and all of the spins can be
characterized by the von-Neumann entropy of the reduced density
matrix of the particle if both the particle and the measuring
apparatus are initially in a pure state. At the time
$t=NT+t^{\prime+}$, the bipartite entanglement can be expressed as
\beqa
S_V(N)&=&-{\mathrm{Tr}}(\rho_{NT+t^{\prime+}}\log_2\rho_{NT+t^{\prime+}})\nonumber\\
&=&-\sum^{\infty}_{i=1}P_i(N)\log_2P_i(N). \eeqa
\begin{figure}
\includegraphics{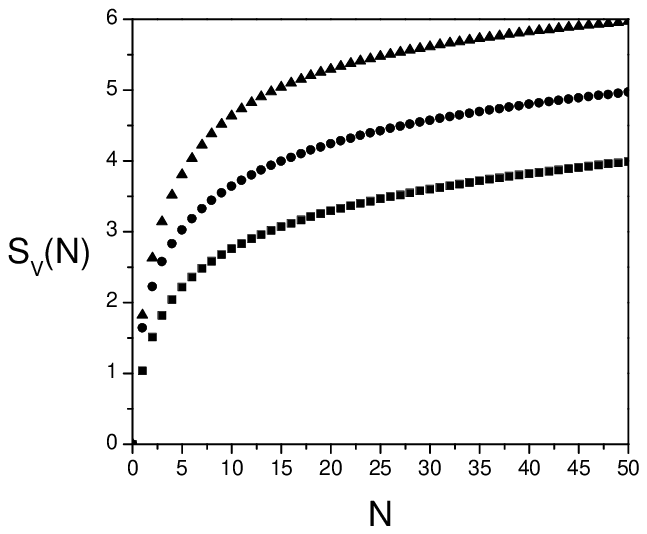}
\caption{ The von-Neumann entropy $S_V(N)$ of the reduced density
matrix of the particle at the time $NT+t^{\prime+}$ are plotted as
a function of $N$ for different values of $R$. The particle is
initially in the ground state and the kicking potential
$V(x)=k\cos(2Rx)$ is chosen as an illustration, $k=\hbar$. (Solid
Square) $R=\pi/4$, (Solid Circle) $R=\pi/2$, (Solid Triangle)
$R=\pi$.}
\end{figure}
In Fig.(4), the von-Neumann entropy $S_V(N)$ in the case with
$V(x)=k\cos(2Rx)$ is plotted for three different values of $R$. It
is shown that the entanglement grows with the kicking steps. The
growth of entanglement exhibits considerable $R$ dependence, and
increasing $R$ to enhance classical diffusion can lead to
significantly larger entanglement for the same field strengths in
this case. By tracing the degrees of freedom of the additional
spins, the particle and the spins $\sigma^{(N)}$ at time
$t=NT+t^{\prime+}$ are in a maximally correlated state
\cite{Rains1999}, which is given by \beqa \rho_{sm}&=&\sum_{n,m}
|n\rangle\langle{n}|U_1\rho_{(N-1)T+t^{\prime+}}U^{\dagger}_1|m\rangle\langle{m}|\nonumber\\
&&\otimes\sigma^{(n,N)}_1\left(\bigotimes_{j}|-\rangle_{(j,N)}\langle-|\right)\sigma^{(m,N)}_1,
\eeqa The partial entanglement quantified by the relative entropy
of entanglement between the particle and the partial spins
$\sigma^{(N)}$ at the time $t=NT+t^{\prime+}$ can be calculated as
\beqa
E_r(N)&=&-\sum^{\infty}_{n=1}P_n(N)\log_2P_n(N)+{\mathrm{Tr}}(\rho_{sm}\log_2\rho_{sm})\nonumber\\
&=&-\sum^{\infty}_{n=1}P_n(N)\log_2P_n(N)\nonumber\\
&&~~~+\sum^{\infty}_{n=1}P_n(N-1)\log_2P_n(N-1). \eeqa Eq.(26)
implies that the partial entanglement between pairs $E_r(N)$
exactly equals $S_V(N)-S_V(N-1)$, i.e., the increment of the
bipartite entanglement. So the bipartite entanglement can also be
rewritten as $S_V(N)=\sum^{N}_{n=1}E_r(n)$, which means that the
bipartite entanglement $S_V(N)$ equals the area of the zone below
the discrete curve of $E_r(n)$ ($n=1,...,N$).
\begin{figure}
\includegraphics{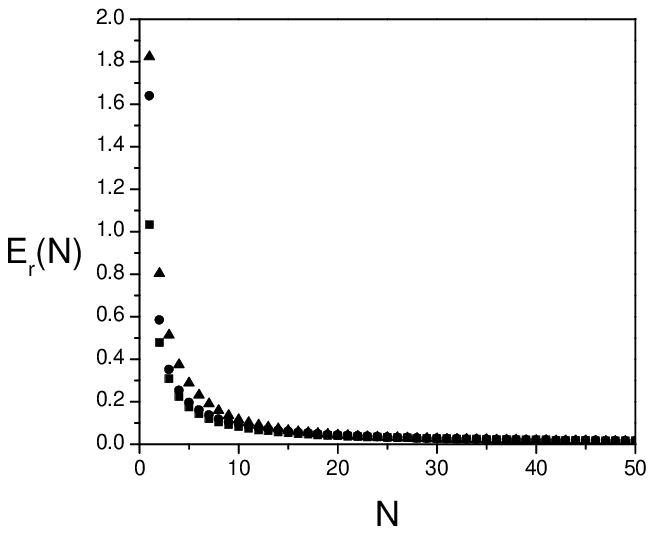}
\caption{ The pairwise entanglement characterized by the relative
entropy of entanglement $E_r(N)$ of the subsystem containing the
particle and the partial spins $\sigma^{(N)}$ at the time
$NT+t^{\prime+}$ are plotted as a function of $N$ for different
values of $R$. The particle is initially in the ground state, and
the spins are initially in the spin down states.
$V(x)=\hbar\cos(2Rx)$, (Solid Square) $R=\pi/4$, (Solid Circle)
$R=\pi/2$, (Solid Triangle) $R=\pi$.}
\end{figure}
In Fig.(5), $E_r(N)$ is displayed for three different values of
$R$. Contrary to the increase of the bipartite entanglement, we
can see that $E_r$ firstly achieves its maximal value at
$t=T+t^{\prime+}$, then rapidly decreases with $N$. There exist
two main factors that could affect the partial entanglement
between pairs. One is the mixedness (ordinarily characterized by
the von-Neumann entropy or the linear entropy), and the other is
the coherence of the particle (characterized by the non-diagonal
elements of density matrix). The increase of coherence can
enhances the entanglement, while the increase of mixedness usually
causes the decrease of entanglement. So, in this case, diminution
of partial entanglement between the particle and the partial spins
$\sigma^{(N)}$ may be owing to the fact that the
measurement-assisted quantum diffusion step by step increase the
mixedness of the system of interest, which plays a more prominent
role in the partial entanglement between pairs than the other
factor.

At the end of this section, some natural questions arise: How does
the bipartite entanglement asymptotically behave? What is the
relation between the asymptotical bipartite entanglement and the
asymptotical diffusion rate? In the present study, these are still
the open questions and need to be investigated in the future work.

\section * {V. CONCLUSIONS}

In this paper, we study the diffusion and entanglement in the
system of a kicked particle in an infinite square well under
frequent measurements. It is shown that the measurement-assisted
diffusion rate can be the constant for a large class of kicking
fields. The asymptotical diffusion rate is also given as
$\frac{1}{2\pi}\int^{\pi}_{0}(V^{\prime})^2dx$. Then we
investigate the bipartite entanglement between the particle and
the whole measuring apparatus and partial entanglement between the
particle and the partial spins. It is found that there exist two
distinct dynamical behaviors of entanglement. The bipartite
entanglement grows with the kicking steps and it gains larger
value for the more chaotic system. However, the partial
entanglement between the system of interest and the partial spins
of the measuring apparatus decreases with the kicking steps. For
the more chaotic system, the bipartite entanglement grows with
higher rate in small time. The increment of the bipartite
entanglement between two coterminous measurements equals to the
partial entanglement between pairs which asymptotically tends to
zero. However, this does not imply that the bipartite entanglement
asymptotically reaches a saturation. Whether there exists a
saturation of bipartite entanglement in such a joint system (in
which the dimension of Hilbert space is infinite) or not is still
a open question in this paper. Nevertheless, we can conclude that
the diffusion of the system and the bipartite entanglement of the
system and the whole measuring apparatus exhibit a cooperation in
the case with the small value of intensity of kicking field,
namely the diffusion enlarges the effective Hilbert space to be
entangled with the measuring apparatus and hence enlarges the
bipartite entanglement, and conversely the bipartite entanglement
destroys the local coherence of the system of interest and
enhances the diffusion.

It has been mentioned that the quantum dot might be a suitable
candidate for experimental realization of the system (1). One of
the key steps is the implementation of the repeat measurements of
the energy. It is unambiguous that direct frequent measurement of
the energy of the atom confined in the quantum dot is not a
trivial task. In the past few years, much attention has been paid
to the quantum measurement schemes in the quantum dots such as the
quantum point contact et al.. In some situations, the quantum
point contact can cause the dephasing of the quantum state. We
briefly discuss the system in the presence of the particular
dephasing which is described by the following master equation \be
\frac{\partial\rho}{\partial{t}}=-\frac{i}{\hbar}[H(t),\rho]-\frac{\gamma{(t)}}{2}[p^2,[p^2,\rho]],\ee
where $H(t)$ is given by the Eq.(1) and $\gamma{(t)}$ is a
constant $\gamma_0$ or
$\sum^{\infty}_{l=-\infty}\varepsilon_0\delta(t-lT-t^{\prime})$.
It is not difficult to verify that two cases are equivalent if
$\gamma_0T=\varepsilon_0$. The evolving density operator in the
latter case is completely reduced to the one described by the
Eq.(2) when $\varepsilon_0\rightarrow\infty$. In the case with
$\gamma_0T\gg1$, the evolution of the density operator described
by Eq.(27) closely approximates to Eq.(2). Therefore, to some
extent, one can also regard the repetitive measurement of energy
as the "noncoherent energy-preserving kick". Though the influence
of the relaxation process on the dynamics need to be taken into
account when the numerical simulation of the realistic situation
is executed. Usually, the dephasing time is much shorter than the
relaxation time in quantum dots. This fact provides us the
possibility for experimentally studying the measurement-induced
dynamical behavior of system (1) via the dephasing process.
Recently, some authors have studied the effects of measurements on
dynamical localization in the kicked rotator model simulated on a
quantum computer \cite{Shep2004}. It was shown that localization
can be preserved for repeated single-qubit measurements. A
transition from a localized to a delocalized phase was revealed,
which depends on the system parameters and the choice of the
measured qubit. To a certain extent, the local measurement on one
qubit can be regarded as a kind of partial dephasing in the kicked
rotator according to the encoding scheme. It could be conjectured
that the different choices of the measured qubit in quantum
computer scheme have the similar effects of different dephasing
coefficients on the dynamical behavior of simulated quantum
chaotic systems. Therefore, it is also very interesting to
investigate whether certain kind of transition can occur in the
system of kicked particle inside an infinite square well when the
dephasing coefficient varies.

Recently, the kicked Bose-Einstein condensate has attracted much
attention. Some authors have investigated the quantum resonance
and anti-resonance of the kicked BEC confined in a one-dimensional
box \cite{Poletti2006}. The results presented in this paper can be
generalized to those systems. It is expected that the frequent
measurement of the energy or dephasing mechanism may significantly
change their dynamical behaviors. If the nonlinear inter-atomic
interaction is ignored, they are reduced to the present case. For
the situation with very small nonlinear inter-atomic interaction,
we can adopt the similar approximation procedure in
Ref.\cite{Poletti2006} to investigate the effect of the nonlinear
inter-atomic interaction on the measurement-induced diffusion.

\section*{ACKNOWLEDGMENT}
This project was supported by the National Natural Science
Foundation of China (Project NO. 10174066).

\bibliographystyle{apsrev}
\bibliography{refmich,refjono}

\end{document}